\documentclass{llncs}
\usepackage[T1]{fontenc}

\usepackage{url}
\usepackage[pdftex]{graphicx}
\usepackage{amsmath,amssymb,xcolor,wrapfig,paralist}
\usepackage{algorithmicx,algorithm}
\usepackage[noend]{algpseudocode}
\usepackage{multirow,colortbl}
\usepackage{bbm}

\newcommand{\goes}[1]{\xrightarrow{#1}}

\usepackage{tikz}
\usetikzlibrary{shapes,positioning,arrows,calc,automata,matrix}
\tikzstyle{state}+=[minimum size = 8mm, inner sep=0,outer sep=1]
\tikzset{->,>=stealth'}

\usepackage[firstpage]{draftwatermark}
\SetWatermarkText{\hspace*{5.5in}\raisebox{4.8in}{\includegraphics[scale=0.1]{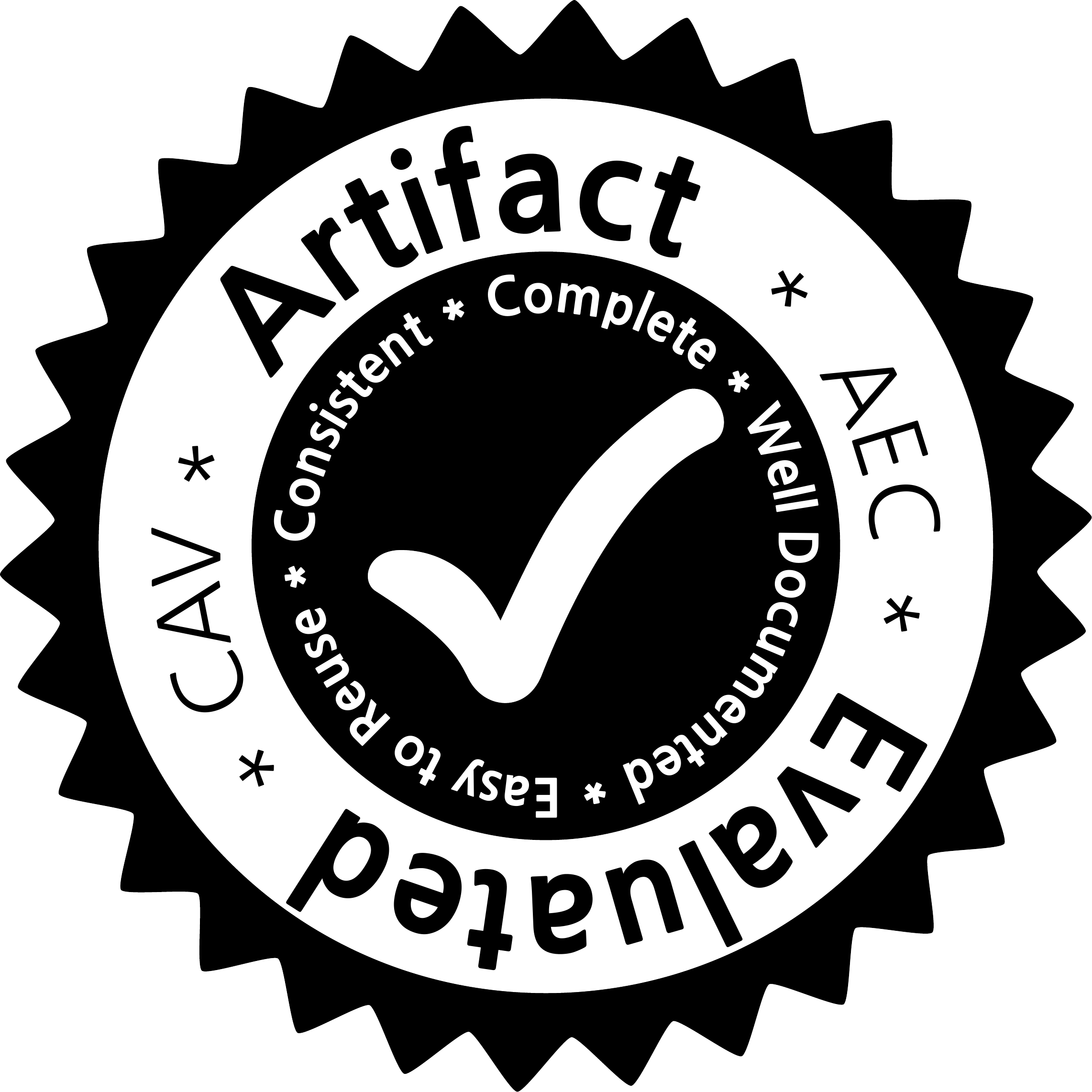}}}
\SetWatermarkAngle{0}

\allowdisplaybreaks

\newcommand{\thmhelperpre}[2]{\newcommand{\theoremlike}[1]{\par\medskip\penalty-250\refstepcounter{theorem}{\bfseries\noindent##1 \ref{#1}.}\itshape}\theoremlike{#2}}
\newcommand{\thmhelperpost}{\par\medskip%
 \renewcommand{\theoremlike}[1]{\par\medskip\penalty-250\refstepcounter{theorem}{\bfseries\noindent##1 \thesection .\thetheorem.}\itshape}%
}

\newcommand{\para}[1]{\smallskip \noindent\textbf{#1.}}

\usepackage{bm}
\renewcommand{\vec}[1]{\bm{#1}}

\usepackage{bbm}

\newcommand{\Nset}{\mathbb N}

\newcommand{\tran}[1]{\xrightarrow{#1}}

\newcommand{\crn}{\mathcal{N}}
\newcommand{\concMC}{\gamma(\crn)}
\newcommand{\absMC}{\alpha(\crn)}
\newcommand{\absIMC}{\alpha(\crn)}

 \usepackage{microtype}
 \newcommand{\myspace}{\vspace*{-0.5em}}
 
 \advance\textwidth6mm 
 \advance\hoffset-3mm
 \advance\textheight6mm
 \advance\voffset-3mm
 
 \let\llncssubparagraph\subparagraph
 \let\subparagraph\paragraph
 \usepackage[compact]{titlesec}
 \let\subparagraph\llncssubparagraph
 
 \titlespacing*{\section}{0pt}{0.7em plus 0.2em minus 0.1em}{0.4em plus 0.05em}
 \titlespacing*{\subsection}{0pt}{0.5em plus 0.1em minus 0.1em}{0.3em plus 0.05em}

\newcommand{\new}[1]{#1}

\title{Semi-Quantitative Abstraction and Analysis of Chemical Reaction Networks\thanks{This work has been supported by  the Czech Science Foundation grant No. GA19-24397S, the IT4Innovations excellence in science project No. LQ1602, and the German Research Foundation (DFG) project KR 4890/2-1 ``Statistical Unbounded Verification''.}\vspace{-.5em}}
\author{Milan \v Ce\v ska\inst1 \and Jan K{{\v r}}et\'insk\'y\inst2 }
\institute{Brno University of Technology, FIT, IT4I Centre of Excellence, Brno, Czech Republic \and Technical University of Munich, Munich, Germany \vspace{-1.5em}}

\def\app{withAppendix}

\begin{document}

\pagestyle{plain}
\maketitle

\begin{abstract}
Analysis of large continuous-time stochastic systems is a computationally intensive task.
In this work we focus on population models arising from chemical reaction networks (CRNs), which play a fundamental role in analysis and design of biochemical systems. 
Many relevant CRNs are particularly challenging for existing techniques due to complex dynamics including stochasticity, stiffness
or multimodal population distributions. 
We propose a novel approach allowing not only to predict, but also to explain both the transient and steady-state behaviour.
It focuses on qualitative description of the behaviour and aims at quantitative precision only in orders of magnitude. 
First we build a compact understandable model, which we then crudely analyse. 
\new{As demonstrated on complex CRNs from literature, our approach reproduces the known results, but in contrast to the state-of-the-art methods, it runs with virtually no computational cost and thus offers unprecedented~scalability.}

%
%
%
	
\end{abstract}

\vspace{-1em}

\section{Introduction}

\new{Chemical Reaction Networks (CRNs) are a versatile language widely used for \emph{modelling and analysis} of biochemical systems~\cite{Chellaboina} as well as for high-level \emph{programming} of molecular devices~\cite{soloveichik2010dna,cardelli2013two}. They provide a compact formalism equivalent to Petri nets \cite{murata1989petri}, Vector Addition Systems (VAS) \cite{karp1969parallel} and distributed population protocols~\cite{angluin2007computational}.
Motivated by numerous potential applications ranging from system biology 
to synthetic biology, 
various techniques allowing simulation and formal analysis of CRNs have been proposed~\cite{gillespie1977exact,salis2005accurate,Verena2013,HybridLNA2016,abate2015adaptive}, and  embodied in the design process of biochemical systems~\cite{giacobbe2015model,heath2008probabilistic,lakin2012design}. The time-evolution of CRNs is governed by the Chemical Master Equation (CME), which describes the probability of the molecular counts of each chemical species. Many important biochemical systems lead to complex dynamics that includes \emph{state space explosion, stochasticity, stiffness,  and multimodality} of the population distributions~\cite{Kampen1992b,goutsias2005quasiequilibrium}, and that fundamentally limits the class of  systems the existing techniques can effectively handle.} More importantly, biologist and  engineers often seek for plausible explanations why the system under study has or has not the required behaviour. In many cases, a set of system simulations/trajectories or population distributions is not sufficient and the ability to provide an accurate explanation for the temporal or steady-state behaviour is another major challenge for the existing techniques.


In order to cope with the computational complexity of the analysis and in order to obtain explanations of the behaviour, we shift the focus from quantitatively precise results to a more qualitative analysis, closer to how a human would behold the system.
Yet we insist on providing at least rough timing information on the behaviour as well as rough classification of probability of different behaviours at the extent of ``very likely'', ``few percent'', ``barely possible'', so that we can conclude on issues such as time to extinction or bimodality of behaviour.
This gives rise to our \emph{semi-quantitative} approach.
We stipulate that analyses in this framework reflect quantities in orders of magnitude, both for time duration and probabilities, but not more than that.
This paradigm shift is reflected on two levels: 
(1)~We abstract systems into semi-quantitative models.
(2)~We analyse systems in a semi-quantitative way.
While each of the two can be combined with a traditional abstraction/analysis, when combined together they provide powerful means to understand systems' behaviour with virtually no computational cost.

\para{Semi-quantitative models}
The states of the models contain information on the current amount of objects of each species as an interval spanning often several orders of magnitude, unless instructed otherwise.
For instance, if an amount of a certain species is to be closely monitored (as a part of the input specification/property of the system) then this abstraction can be finer.
Similarly, whenever the analysis of a previous version of the abstraction points to the lack of precision in certain states, preventing us to conclude which of the possible behaviours is prevalent, the corresponding refinement can take place.
Further, the rates of the transitions are also captured only with such imprecision.
The crucial point allowing for existence of such models that are small, yet faithful, is our concept of \emph{acceleration}. 
It captures certain \emph{sequences} of transitions.
It eliminates most of the non-determinism that paralyses other types of abstractions, which are too over-approximative, unable to conclude anything, but safety properties.

\para{Semi-quantitative analysis}
Instead of performing exact transient or steady-state analysis, we can consider most probable transitions and then carefully lift this to most probable temporal behaviours.
Technically, this is done by \emph{alternating between transient and steady-state analysis} where only some rates and transitions are taken into account at different stages.
In order to further facilitate the resulting insight of the human on the result of the analysis, we provide an algorithm to perform this analysis with virtually no computation effort and thus possibly manually.
The trivial computations immediately pinpoint why certain behaviours occur.
Moreover, less likely behaviours can also be identified easily, to any desired degree of improbability (dozens of percent, promilles etc.).

To summarise, the first step yields tiny models, allowing for a synoptic observation of the model; 
due to their size these models can be either analysed easily using standard means, or can be subject to the second step.
The second step provides an efficient approximative analysis, which is also very illustrative due to the limited use of quantities.
It can be applied to any system; however, it is particularly interesting in connection with the models coming from the first step since (i) no extra effort (size, computation) is wasted on overly precise treatment that is ignored by the other step, and (ii) together they yield an understandable explanation of the behaviour.
An entertaining feature of this paradigm is that the stiffer (with rates at hugely different time scales) the system is the easier it is to analyse.

To demonstrate the capabilities of our approach, we consider three challenging and biologically relevant case studies that have been used in literature to evaluate state-of-the-art methods for the CRN analysis. It has been shown that  many approaches fail, either due to time-outs or incapability to capture differences in behaviours, and some tailored ones require considerable computational effort, e.g. an hour of computation. Our experiments clearly show that the proposed approach can deliver results that yield qualitatively same information, more understanding and can be computed in minutes by hand (or within a fraction of a second by computer).

\smallskip \noindent\textbf{Our contribution} can be summarized as follows: \vspace*{-.5em}
\begin{itemize}
	\item We propose a novel \emph{semi-quantitative} framework for analysis of CRN and similar population models, focusing on explainability of the results and low complexity, with quantitative precision limited to orders of magnitude.
	\item \new{An algorithm for abstracting CRNs into semi-quantitative models  based on interval abstraction of the species population and on transition acceleration.}
	\item An algorithm for semi-quantitative analysis that replaces exact numerical computation by exploring the most probable transitions and alternating transient and steady-state analysis.
	\item \new{We consider three challenging CRNs thoroughly studied in literature and demonstrate that the semi-quantitative abstraction and analysis gives us a unique tool that is able to accurately predict and explain both transient and steady-state behaviour of complex CRNs in a fraction of a second.}

\end{itemize}

\subsection*{Related Work}
To the best of our knowledge, there does not exist any  abstraction  of CRNs similar to the proposed approach. Indeed, there exist various abstraction and approximation schemes for CRNs that improve the performance and scalability  of both the simulation-based and the numerical-based techniques.
In the following paragraphs, we discuss the most relevant  directions and the links to our approach. 

\para{Approximate semantics for CRNs}
For CRNs including  large populations of species, 
fluid (mean-field) approximation techniques can be applied~\cite{Bortolussi2012}
and extended to approximate higher-order moments~\cite{Engblom06}: 
these deterministic approximations lead to a set of ordinary differential equations (ODEs). An alternative is to approximate the CME as a continuous-state stochastic process. The Linear Noise Approximation (LNA) is a Gaussian process which has been derived as an approximation of the CME~\cite{Kampen1992b,ethier2009markov} and describes the time evolution of expectation and variance of the species in terms of ODEs. Recently, an aggregation scheme over ODEs that aims at understanding the dynamics of large CRNs has been proposed in~\cite{cardelli2017maximal}. In contrast to our approach, the deterministic approximations cannot  adequately capture the stochasticity of CRNs
caused by low population species.

To mitigate this drawback, various \emph{hybrid models} have been proposed. The common idea of these models is as follows: the dynamics of low population species is described by the discrete stochastic process and the dynamics of large population species is approximated by a continuous process. The particular hybrid models differ in the approximation of the large population species. In~\cite{henzinger2010hybrid}, a pure deterministic semantics for large population species is used. The moment-based description for medium/high-copy number species was used in~\cite{Verena2013}. The LNA approximation and an adaptive partitioning of the species according to leap conditions (that is more general than partitioning based on population thresholds) was proposed in~\cite{HybridLNA2016}.  All hybrid models have  to deal with interactions between low and large population species. In particular, the dynamics of the stochastic process describing the low-population species is conditioned by the continuous-state describing the concentration of the large-population species. \new{The numerical analysis of such conditioned stochastic process is typically  a computationally  demanding task that limits the scalability.}

In contrast, \new{our approach does not explicitly partition the species, but rather abstracts the concrete species population using an interval abstraction and tries to effectively capture both the stochastic and the deterministic behaviour with the help of the accelerated transitions. As we already emphasised, the proposed abstraction and analysis avoids any numerical computation of precise quantities.}

%


\para{Reduction techniques for stochastic  models}
A widely studied reduction method for Markov models is state aggregation based on lumping \cite{B99} or \mbox{(bi-)simu}\-lation equivalence \cite{BK08}, with
the latter notion in its exact \cite{Larsen94} or approximate \cite{DLT08} form.
Approximate notions of equivalence
have led to new abstraction/refinement techniques for the numerical verification of Markov models
over finite \cite{DAK12} as well as uncountably-infinite state spaces \cite{AKLP10,SA13,SA14}.
\new{Several approximate aggregation schemes leveraging the structural properties of CRNs were proposed~\cite{Madsen2012,Zhang09,ferm2009adaptive}.
Abate et al. proposed an adaptive aggregation that gives formal guarantees on the approximation error, but typically provide lower state space reductions~\cite{abate2015adaptive}. } 
\new{Our approach shares the idea of abstracting the state space by aggre\-gating some states together. Similarly to~\cite{Madsen2012,Zhang09,ferm2009adaptive} we partition the state space based on the species population, i.e. we also introduce the population levels. In contrast to the aforementioned aggregation schemes, we propose a novel abstraction of the transition relation based on the acceleration. It allows us to avoid the numerical solution of the approximate CME and thus achieve a better 
reduction while providing  an accurate predication of the system behaviour.}

\new{Alternative methods to deal with large/infinite state spaces are based on a state truncation trying to eliminate insignificant states, i.e., states reached only with a negligible probability. These methods, including  finite state projections~\cite{munsky2006finite}, 
sliding window abstractions~\cite{Henzinger2009}, or fast adaptive uniformisation ~\cite{Mateescu10}, are able to quantify the total probability mass that is lost due to the truncation, but typically cannot effectively handle systems involving a stiff behaviour and multimodality~\cite{HybridLNA2016}.}



\para{Simulation-based analysis}
Transient analysis of CRNs can be performed using the Stochastic Simulation Algorithm (SSA)~\cite{gillespie1977exact}. Note that the SSA produces a single realisation of the stochastic process, whereas the stochastic solution of CME gives the probability distribution of each species over time.  Although 
simulation-based analysis is generally  faster than direct solution of the stochastic process underlying the given CRN, obtaining good accuracy necessitates potentially large numbers of simulations and can be very time consuming. 

Various partitioning schemes for species and reactions have been proposed for the purpose of speeding up the SSA in multi-scale systems~\cite{goutsias2005quasiequilibrium,rao2003stochastic,salis2005accurate}. For instance, Yao et al. introduced the slow-scale SSA~\cite{cao2005slow}, where they distinguish between fast and slow species. Fast species are then treated assuming they reach equilibrium much faster than the slow ones. Adaptive partitioning of the species has been considered in~\cite{ganguly2015jump,hepp2015adaptive}. \new{In contrast to the simulation-based analysis, our approach (i) provides a compact explanation of the system behaviour in the form of tiny models allowing for a synoptic observation and (ii) can easily reveal less probable behaviours.}

\section{Chemical Reaction Networks}
In this paper, we assume familiarity with standard verification of (continuous-time) probabilistic systems, e.g. \cite{BK08}.
For more detail, see \cite[Appendix]{techreport}.
\paragraph{CRN Syntax.}
A \emph{chemical reaction network (CRN)} $\crn=(\Lambda,\mathcal{R})$ is a pair of finite sets, where $\Lambda$ is a set of \emph{species}, $|\Lambda|$ denotes its size, and $\mathcal{R}$ is a set of reactions. Species in $\Lambda$ interact according to the reactions in $\mathcal{R}$. A \emph{reaction} $\tau \in \mathcal{R}$ is a triple $\tau=(r_{\tau},p_{\tau},k_{\tau})$, where $r_{\tau} \in  \mathbb{N}^{|\Lambda|}$ is the \emph{reactant complex}, 
$p_{\tau} \in  \mathbb{N}^{|\Lambda|}$ is the \emph{product complex} and $k_{\tau} \in \mathbb{R}_{>0} $ is the coefficient associated with the rate of the reaction. $r_{\tau}$ and $p_{\tau}$ represent the stoichiometry of reactants and products.
Given a reaction $\tau_1=(  [1,1,0],[0,0,2],k_1 )$, we often refer to it as $\tau_1 : \lambda_1 + \lambda_2 \, \goes{k_1}  \,    2\lambda_3 $. 
\vspace{-0.3em}
\paragraph{CRN semantics.}\label{Concrete Semantics}
Under the usual assumption of mass action kinetics, the \emph{stochastic} semantics of a CRN $\crn$ is generally given in terms of a discrete-state, continu\-ous-time stochastic process $\mathbf{X(t)}=(X_1(t),X_2(t), \ldots, X_{|\Lambda|}(t) ,t\geq 0)$ \cite{ethier2009markov}. The \emph{state change} associated to the reaction $\tau$ is defined by $\upsilon_{\tau}=p_{\tau} - r_{\tau}$, i.e. the state $\mathbf{X}$ is changed 
 to $\mathbf{X}' = \mathbf{X} + \upsilon_{\tau}$, which we denote as
$\mathbf{X} \goes{\tau}  \mathbf{X}'$. For example, for $\tau_1$ as above, we have $\upsilon_{\tau_1}=[-1,-1,2]$. 
For a reaction to happen in a state $\mathbf{X}$, all reactants have to be in sufficient numbers. The \emph{reachable
state space} of~$\mathbf{X(t)}$, denoted as~$\mathbf{S}$, is the
set of all states reachable by a  sequence of reactions from a given \emph{initial state} $\mathbf{X}_0$. 
The set of reactions changing the state $\mathbf{X}_i$ to the
state $\mathbf{X}_j$ is denoted as $\mathsf{reac}(\mathbf{X}_i,\mathbf{X}_j) =
\{\tau \mid \mathbf{X}_i \goes{\tau} \mathbf{X}_j \} $.

The behaviour of the
stochastic system $\mathbf{X(t)}$ can be described by the (possibly infinite) continuous-time Markov
chain (CTMC) $\concMC = (\mathbf{S}, \mathbf{X}_0, \mathbf{R})$ where the
transition matrix $\mathbf{R}(i,j)$ gives the probability of a transition from
$\mathbf{X}_i$ to $\mathbf{X}_j$. Formally, 
\vspace{-.5em}
\begin{equation}
 \mathbf{R}(i,j) = \sum_{\tau \in
  \mathsf{reac}(\mathbf{X}_i,\mathbf{X}_j) } k_\tau \cdot C_{\tau,i} \quad  \mbox{where} \quad C_{\tau,i} =
\prod_{\ell = 1}^{N}\binom{\mathbf{X}_{i,\ell}}{r_\ell}
\label{eq:rate}\tag{R}
\end{equation} 
corresponds to the population dependent term of the \emph{propensity function}
where $\mathbf{X}_{i,\ell}$ is $\ell$th component of the state $\mathbf{X}_{i}$ and $r_\ell$ is the stoichiometric coefficient of the $\ell$-th reactant in the reaction $\tau$.
The CTMC $\concMC$ is the accurate representation of CRN $\crn$, but---even when finite---not scalable in practice because of the state space explosion problem \cite{kwiatkowska2014probabilistic,heath2008probabilistic}.

%


%
%

\section{Semi-quantitative Abstraction}

In this section, we describe our abstraction.
We derive the desired CTMC conceptually in several steps, which we describe explicitly, although we implement the construction of the final system directly from the initial CRN.

\subsection{Over-approximation by Interval Abstraction and Acceleration}
Given a CRN $\crn=(\Lambda,\mathcal R)$, we first consider an interval continuous-time Markov decision process (interval CTMDP\footnote{Interval CTMDP is a CTMDP with lower/upper bounds on rates. Since it serves only as an intermediate formalism to ease the presentation, we refrain from formalising it here.}), which is a finite abstraction of the infinite~$\concMC$.
Intuitively, abstract states are given by intervals on sizes of populations with an additional specific that the abstraction captures enabledness of reactions. 
The transition structure follows the ideas of the standard may abstraction and of the three-valued abstraction of continuous-time systems \cite{DBLP:conf/cav/KatoenKLW07}.
A technical difference in the latter point is that we abstract rates into intervals instead of uniformising the chain and then only abstracting transition probabilities into intervals; this is necessary in later stages of the process.
The main difference is that we also treat certain sequences of actions, which we call acceleration.

\para{Abstract domains}
The first step is to define the abstract domain for the population sizes.
For every species $\lambda\in\Lambda$, we define a finite partitioning $A_\lambda$ of $\Nset$ into intervals, reflecting the rough size of the population.
Moreover, we want the abstraction to reflect whether a reaction is enabled.
Hence we require that $\{0\}\in A_\lambda$ for the case when the coefficients of this species as a reactant is always $0$ or $1$;
in general, for every $i<\max_{\tau\in\mathcal R}r_\tau(\lambda)$ we require $\{i\}\in A_\lambda$.

The abstraction $\alpha_\lambda(n)$ of a number $n$ of a species $\lambda$ is then the $I\in A_\lambda$ for which $n\in I$.
The state space of $\absIMC$ is the product $\prod_{\lambda\in\Lambda}A_\lambda$ of the abstract domains with the point-wise defined abstraction $\alpha(\vec n)_\lambda=\alpha_\lambda(\vec n_\lambda)$.

The abstract domain for the rates according to (\ref{eq:rate}) is the set of all real intervals.

Transitions from an abstract state are defined as the may abstraction as follows.
Since our abstraction reflect enabledness, the same set of action is enabled in all concrete states of a given abstract state.
The targets of the action in the abstract setting are abstractions of all possible concrete successors, i.e. 
$\mathit{succ}(s,a):=\{\alpha(\vec n)\mid \vec m\in s, \vec m\tran{a}\vec n\}$,
in other words, the transitions enabled in at least one of the respective concrete states.
The abstract rate is the smallest interval including all the concrete rates of the respective concrete transitions.
This can be easily computed by the corner-points abstraction (evaluating only the extremum values for each species) since the stoichiometry of the rates is monotone in the population sizes.

\para{High-level of non-determinism}
The (more or less) standard style of the abstraction above has several drawbacks---mostly related to the high degree of non-determinism for rates---which we will subsequently discuss.

Firstly, in connection with the abstract population sizes, transitions to different sizes only happen non-deterministically, leaving us unable to determine which behaviour is probable. For example, consider the simple system given by $\lambda\tran{d}\emptyset$ with $k_d=10^{-4}$ so the degradation happens on average each $10^4$ seconds.
Assume population discretisation into $[0],[1..5],[6..20],[21..\infty)$ with abstraction depicted in Fig.~\ref{fig:acc}.
While the original system obviously moves from $[6..20]$ to $[1..5]$ very probably in less than $15\cdot 10^{4}$ seconds, the abstraction cannot even say that it happens, not to speak of estimating the time.

\tikzset{
	state/.style={
		rectangle,
		rounded corners,
		draw=black,
		minimum height=2em,
		minimum width=3em,
		inner sep=4pt,
		text centered,
	}
}
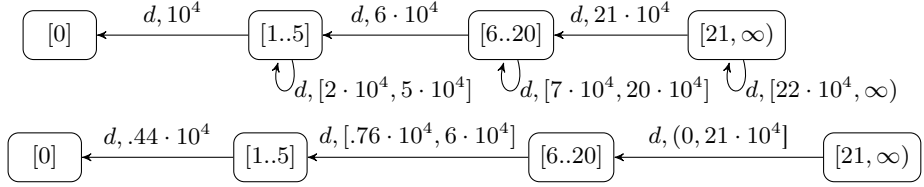
\begin{figure}[t]
	\myspace
	\centering
\begin{tikzpicture}
\node[state](s0) at(0,0) {$[0]$};
\node[state](s1) at(3,0) {$[1..5]$};
\node[state](s6) at(6,0) {$[6..20]$};
\node[state](s21) at(9,0) {$[21,\infty)$};

\draw[->] 
(s1) edge node[above]{$d,10^4$} (s0)
(s6) edge node[above]{$d,6\cdot10^4$} (s1)
(s21) edge node[above]{$d,21\cdot10^4$} (s6)
(s1) edge[loop below] node[right] {$d,[2\cdot10^4,5\cdot10^4]$} ()
(s6) edge[loop below] node[right] {$d,[7\cdot10^4,20\cdot10^4]$} ()
(s21) edge[loop below] node[right] {$d,[22\cdot10^4,\infty)$} ()
;
\end{tikzpicture}

\begin{tikzpicture}
\node[state](s0) at(0,0) {$[0]$};
\node[state](s1) at(3,0) {$[1..5]$};
\node[state](s6) at(7,0) {$[6..20]$};
\node[state](s21) at(11,0) {$[21,\infty)$};

\draw[->] 
(s1) edge node[above]{$d,.44\cdot10^4$} (s0)
(s6) edge node[above]{$d,[.76\cdot10^4,6\cdot10^4]$} (s1)
(s21) edge node[above]{$d,(0,21\cdot10^4$]} (s6)
;
\end{tikzpicture}
\caption{Above: Interval CTMDP abstraction with intervals on rates and non-determinism. 
	Below: Interval CTMC abstraction arising from acceleration.}
\label{fig:acc}
\end{figure}

\para{Acceleration} 
To address this issue, we drop the non-deterministic self-loops and transitions to higher/lower populations in the abstract system.\footnote{One can also preserve the non-determinism for the special case when one of the transitions leads to a state where some action ceases to be enabled. While this adds more precision, the non-determinism in the abstraction makes it less convenient to handle.} 
Instead, we \emph{``accelerate''} their effect: 
We consider sequences of these actions that in the concrete system have the effect of changing the population level.
In our example above, we need to take the transition 1 to 13 times from $[6..20]$ with various rates depending on the current concrete population, in order to get to $[1..5]$.
This makes the precise timing more complicated to compute.
Nevertheless, the expected time can be approximated easily: here it ranges from $\frac 1 6\cdot 10^{4}=0.17\cdot 10^4$ (for population 6) to roughly $(\frac1{20}+\frac1{19}+\cdots+\frac 1{6})\cdot 10^{4}=1.3\cdot 10^4$ (for population~20).
This results in an interval~CTMC.\footnote{The waiting times are not distributed according to the rates in the intervals. It is only the expected waiting time (reciprocal of the rate) that is preserved. Nevertheless, for ease of exposition, instead of labelling the transitions with expected waiting times we stick to the CTMC style with the reciprocals and formally treat it as if the label was a real rate.}


\para{Concurrency in acceleration}
The accelerated transitions can due to higher number of occurrences be considered continuous or deterministic, as opposed to discrete stochastic changes as distinguished in the hybrid approach.
The usual differential equation approach would also take into account other reactions that are modelled deterministically and would combine their effect into one equation.
In order to simplify the exposition and computation and---as we see later---without much loss of precision, we can consider only the fastest change (or non-deterministically more of them if their rates are similar).\footnote{Typically the classical concurrency diamond appears and the effect of the other accelerated reactions happen just after the first one.}

\subsection{Operational Semantics: Concretisation to a Representative}

The next disadvantage of classical abstraction philosophy, manifested in the interval CTMC above is that the precise-valued intervals on rates imply high computational effort during the analysis.
Although the system is smaller, standard transient analysis is still quite expensive.

\para{Concretisation}
In order to deal with this issue, the interval can be approximated roughly by the expected time it would take for an average population in the considered range, in our example the ``average'' representative is 13.
Then the first transition occurs with rate $13\cdot 10^{-4}=10^{-3}$ and needs to happen 7 times, yielding expected time $7/13\cdot 10^4=0.5\cdot 10^4$ (ignoring even the precise slow downs in the rates as the population decreases).
Already this very rough computation yields relative precision with factor 3 for all the populations in this interval, thus yielding the correct order of magnitude with virtually no effort.
We lift the concretisation naturally to states and denote the concretisation  of abstract state $s$ by $\gamma(s)$.
The complete procedure is depicted in Algorithm~\ref{alg:general}.

The concretisation is one of the main points where we deliberately drop a lot of quantitative information, while still preserving some to conclude on big quantitative differences.
Of course, the precision improves with more precise abstract domains and also with higher differences on the original rates.

\begin{algorithm}[t]
	\caption{Semi-quantitative abstraction CTMC $\absMC$}\label{alg:general}
	\begin{algorithmic}[1]
		\State $A\gets\prod_{\lambda\in \Lambda} A_\lambda$ \Comment{States}
		\For{$\vec a\in A$} \Comment{Transitions}
		\State $\vec c\gets\gamma(\vec a)$ \Comment{Concrete representative}
		\For{each $\tau$ enabled in $\vec c$} 
		\State $r\gets$rate of $\tau$ in $\vec c$ \Comment{According to (\ref{eq:rate})}
		\State $\vec a'\gets\alpha(\vec c + \upsilon_\tau)$ \Comment{Successor} 
		\State set $\vec a\xrightarrow{\tau}\vec a'$ with rate $r$ 
		\EndFor
		\For{self-loop $\vec a\xrightarrow{\tau}\vec a$}  \Comment{Accelerate self-loops}
		\State $n_\tau\gets\min\{n\mid \alpha(\vec c + n\cdot\upsilon_\tau)\neq \vec a\}$ \Comment{the number  of $\tau$ to change the abstract state} 
		\State $\vec a'\gets \alpha(\vec c + n_\tau\cdot\upsilon_\tau)$ \Comment{Acceleration successor}
		\State instead of the self-loop with rate $r$, set $\vec a\xrightarrow{\tau}\vec a'$ with rate $n_\tau\cdot r$ 
		\EndFor	
		\EndFor			
	\end{algorithmic}
\end{algorithm}	

It remains to determine the representative for the unbounded interval.
In order to avoid infinity, we require an additional input for the analysis, which are deemed upper bounds on possible population of each species. 
In cases when any upper bound is hard to assume, we can analyse the system with a random one and see if the last interval is reachable with significant probability.
If yes, then we need to use this upper bound as a new point in the interval partitioning and try a higher upper bound next time.
In general, such conditions can be checked in the abstraction and their violation implies a recommendation to refine the abstract domains accordingly.

\para{Orders-of-magnitude abstraction}
Such an approximation is thus sufficient to determine most of the time whether the acceleration (sequence of actions) happens sooner or later than e.g. another reaction with rate $10^{-6}$ or $10^{-2}$.
Note that this \emph{decision} gets more precise not only as we refine the population levels, but also as the system gets stiffer (the concrete values of the rates differ more), which are normally harder to analyse.
For the ease of presentation in our case studies, we shall depict only the magnitude of the rates, i.e. the decadic logarithm rounded to an integer.

\para{Non-determinism and refinement}
If two rates are close to each other, say of the same magnitude (or difference 1), such a rough computation (and rough population discretisation) is not precise enough to determine which of the reactions happens with high probability sooner.
Both may be happening roughly at the same pace, or with more information we could conclude one of them is considerably faster.
This introduces an uncertainty, showing different behaviours are possible depending on the exact quantities.
This indicates points where refinement might be needed if more precise results are required. 
For instance, with rates of magnitudes 2 and 3, the latter should be happing most of the time, the former only with a few percent chance.
If we want to know whether it is rather tens of percent or tenths of percent, we should refine the abstraction.

\enlargethispage{\baselineskip}


%

\section{Semi-quantitative Analysis}

In this section, we present an approximative analysis technique that describes the most probable transient and steady-state behaviour of the system (also with rough timing) and on demand also the (one or more orders of magnitude) less probable behaviours.
As such it is robust in the sense that it is well suited to work with imprecise rates and populations.
It is computationally easy (can be done in hand in time required for a computer by other methods), while still yielding significant quantitative results (``in orders of magnitude'').
It does not provide exact error guarantees since computing them would be almost as expensive as the classical analysis.
It only features trivial limit-style bounds: if the population abstraction gets more and more refined, the probabilities converge to those of the original system; further, the higher the separation between the rate magnitudes, the more precise the approximation is since the other factors (and thus the incurred imprecisions) play less significant role.

Intuitively, the main idea---similar to some multi-rate simulation techniques for stiff systems---is to ``simulate'' ``fast'' reactions until the steady state and then examine which slower reactions take place.
However, ``fast'' does not mean faster than some constant, but faster than other transitions in a given state.
In other words, we are not distinguishing fast and slow reactions, but tailor this to each state separately.
Further, ``simulation'' is not really a stochastic simulation, but a deterministic choice of the fastest available transition.
If a transition is significantly faster than others then this yields what a simulation would yield.
When there are transitions with similar rates, e.g.\ with at most one order of magnitude difference, then both are taken into account as described in the following definition.

\para{Pruned system} 
Consider the underlying graph of the given CTMC.
If we keep only the outgoing transitions with the maximum rate in each state, we call the result \emph{pruned}.
If there is always (at most) one transition then the graph consists of several paths leading to cycles.
In general when more transitions are kept, it has bottom strongly connected components (bottom SCCs, BSCCs) and some transient parts.

We generalise this concept to \emph{$n$-pruning} that preserves all transitions with a rate that is not more than $n$ orders of magnitude smaller than the maximum rate in the state.
Then the pruning above is $0$-pruning, $1$-pruning preserves also transitions happening up to 10 times slower, which can thus still happen with dozens of percent, $2$-pruning is relevant for analysis where behaviour occurring with units of percent is also tracked etc.

\para{Algorithm idea}
Here we explain the idea of Algorithm~\ref{alg:analyse}.
The transient parts of the pruned system describe the most probable behaviour from each state until the point where visited states start to repeat a lot (steady state of the pruned system). 
In the original system, the usual behaviour is then to stay in this SCC $C$ until one of the pruned (slower) reactions occurs, say from state $s$ to state $t$.
This may bring us to a different component of the pruned graph and the analysis process repeats.
However, $t$ may also bring us back into $C$, in which case we stay in the steady-state, which is basically the same as without the transition from $s$ to $t$.
Further, $t$ might be in the transient part leading to $C$, in which case these states are added to $C$ and the steady state changes a bit, spreading the distribution slightly also to the previously transient states.
Finally, $t$ might be leading us into a component $D$ where this run was previous to visiting $C$.
In that case, the steady-state distribution spreads over all the components visited between $D$ and $C$, putting a probability mass to each with a different order of magnitude depending on all the (magnitudes of) sojourn times in the transient and steady-state phases on the way.


\begin{algorithm}[t]
	\caption{Semi-quantitative analysis}\label{alg:analyse}
	
	\scalebox{0.99}{
	\begin{minipage}{1.0\linewidth}
	\begin{algorithmic}[1]
		\State $W\gets\emptyset$ \Comment{worklist of SCCs to process}
		\State add $\{\text{initial state}\}$ to $W$ and assign iteration 0 to it \Comment{artificial SCC to start the process}
		\While{$W\neq\emptyset$}	
			\State $C\gets$pop $W$
			\Statex\Comment{Compute and output  steady state or its approximation}
			\State steady-state of $C$ is approximately $\mathit{minStayingRate}/(m\cdot\mathit{stayingRate}(\cdot))$ \label{l:ss}
			\If{$C$ has no exits} continue \Comment{definitely bottom SCC, final steady state}
			\EndIf
			\Statex\Comment{Compute and output exiting transitions and the time spent in $C$}
			\State $\mathit{exitStates}\gets\arg\min_C (\mathit{stayingRate}(\cdot) / \mathit{exitingRate}(\cdot))$ \Comment{Probable exit points}\label{l:exits}
			\State $\mathit{minStayingRate}\gets$minimum rate in $C$, $m\gets$\#occurrences there
			\State $\mathit{timeToExit}\gets\mathit{stayingRate}(s) \cdot m /(|\mathit{exitStates}|\cdot\mathit{minStayingRate} \cdot \mathit{exitingRate}(s))$ 
			\Statex \hspace*{8cm}for (arbitrary) $s\in\mathit{exitStates}$ \label{l:ssex}
			\ForAll{$s\in\mathit{exitsStates}$} \Comment{Transient analysis}
			\State $t\gets$target of the exiting transition
			\State $T\gets$SCCs reachable in the pruned graph from $t$
			\State thereby newly reached transitions get assigned iteration of $C$ + 1
			\For{$D\in T$}
			\Statex\Comment{Compute and output time to get from $t$ to $D$}
			\State $\mathit{minRate}\gets$minimum rate on the way from $t$ to $D$, $m\gets$\#occurrences there
			\State $\mathit{transTime}\gets m/\mathit{minRate}$\label{l:trans}
			\Statex\Comment{Determine the new SCC}
			\If{$D=C$} \Comment{back to the current SCC}
			\State add to $W$ the union of $C$ and the new transient path $\tau$ from $t$ to $C$
			\State in later steady-state computation, the states of $\tau$ will have probability \Statex\hspace*{5cm}  smaller by a factor of $\mathit{stayingRate}(s) / \mathit{exitingRate}(s)$ 
			\ElsIf {$D$ was previously visited} \Comment{alternating between different SCCs}
			\State add to $W$ the merge of all SCCs visited between $D$ and $C$ (inclusively)
			\State in later steady-state computation, reflect all $\mathit{timeToExit}$ and $\mathit{transTime}$ \Statex\hfill between $D$ and $C$ 
			\Else
			\Comment{new SCC}
			\State add $D$ to $W$
			\EndIf
			\EndFor		
			\EndFor
		\EndWhile
	\end{algorithmic}
\end{minipage}
}
~\\
MACROS:\\
$\mathit{stayingRate}(s)$ is the rate of transitions from $s$ in the pruned graph\\
$\mathit{exitingRate}(s)$ is the maximum rate of transitions from $s$ not in the pruned graph
\end{algorithm}

Using the macros defined in the algorithm, the correctness of the computations can be shown as follows.
For the time spent in the transient phase (line \ref{l:trans}), we consider the slowest sojourn time on the way times the number of such transitions; this is accurate since the other times are by order(s) of magnitude shorter, hence negligible.
The steady-state distribution on a BSCC of the pruned graph can be approximated by the $\mathit{minStayingRate}/(m\cdot\mathit{stayingRate}(\cdot))$ on line \ref{l:ss}. Indeed, it corresponds to the steady-state distribution if the BSCC is a cycle and the $\mathit{minStayingRate}$ significantly larger than other rates in the BSCC since then the return time for the states is approximately $m/\mathit{minStayingRate}$ and the sojourn time $1/\mathit{stayingRate}(\cdot)$.
The component is exited from $s$ with the proportion given by its steady-state distribution times the probability to take the exit during that time.
The former is approximated above; the latter can be approximated by the density in $0$, i.e. by $\mathit{exitingRate}(s)$, since the staying rate is significantly faster.
Hence the candidates for exiting are maximising $\mathit{exitingRate}(\cdot)/\mathit{stayingRate}(\cdot)$ as on line \ref{l:exits}.
There are $|\mathit{exitStates}|$ candidates for exit and the time to exit the component by a particular candidate $s$ is the expected number of visits before exit, i.e. $\mathit{stayingRate}(s)\cdot\mathit{exitingRate}(s)$ times the return time $m\cdot\mathit{minStayingRate}$, hence the expression on line \ref{l:ssex}.

\begin{figure}[t!]
	\centering
	\begin{tikzpicture}
	\node[state,initial above,initial text=](s0) at(0,0) {$s_0$};
	\node[state](s1) at(2,0) {$s_1$};
	\node[state](s2) at(4,0) {$s_2$};
	\node[state](s3) at(6,0) {$s_3$};
	\node[state](t) at(-2,0) {$t$};
	\node[state](u) at(8,0) {$u$};
	
	\draw[->] 
	(s0) edge node[above]{$1$} (s1)
	(s0) edge node[above]{$1$} (t)
	(s1) edge node[above]{$10$} (s2)
	(s2) edge node[above]{$10$} (s3)
	(s3) edge[bend right=25pt] node[above]{$100$} (s1)
	(t) edge[loop above] node[left]{$1$} ()
	(u) edge[loop above] node[right]{$1$} ()	
	(s3) edge[dotted] node[above]{$100$} (u)
	(s2) edge [bend left,dashed] node[below]{$1$} (s0)
	(s3) edge [bend left,dashed] node[below]{$10$} (s1)	
	;
	\end{tikzpicture}
	\caption{Alternating transient and steady-state analysis.}
	\label{fig:analysis}
\end{figure}
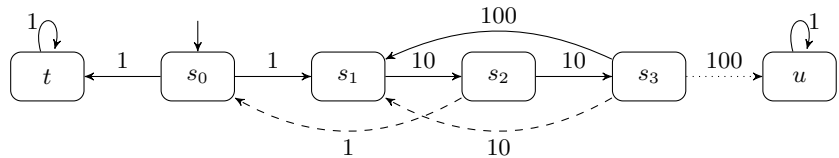

For example, consider the system in Fig.~\ref{fig:analysis}.
Iteration 1 reveals the part with solid lines with two (temporary) BSCCs $\{t\}$ and $\{s_1,s_2,s_3\}$.
The former turns out definitely bottom.
The latter has a steady state proportional to $(10^{-1},10^{-1},100^{-1})$. 
Its most probable exits are the dashed ones, identified in the subsequent iteration~2, probable proportionally to $(1/10,10/100)$; the expected time to take them is $10\cdot2/(2\cdot10\cdot1)=1=100\cdot2/(2\cdot10\cdot10)$.
The latter leads back to the current SCC and does not change the set of BSCCs (hence in our examples below we often either skip or merge such iterations for the sake of readability).
In contrast, the former leads to a previous SCC; thereafter $\{s_1,s_2,s_3\}$ is no more a bottom SCC and consequently the third exit to $u$ is not even analysed.
Nevertheless, it could still happen with minor probability, which can be seen if we consider 1-pruning instead.

%
%
%
%
%
%
%
%
%
%
%
%
%
%
%
%


%
%

\section{Experimental Evaluation and Discussion} 

In order to demonstrate the applicability and accuracy of our approach, we selected the following three biologically relevant case studies. 
(1) stochastic model of gene expression~\cite{golding2005real,Verena2013}, (2) Goutsias's model~\cite{goutsias2005quasiequilibrium} describing transcription regulation of a repressor protein in bacteriophage $\lambda$
and (3) viral infection model~\cite{srivastava2002stochastic}. 

Although the underlying CRNs are quite small (up to 5 species and 10 reaction), their analysis is very challenging: (i) the stochasticity has a strong impact on the dynamics of these systems and thus purely deterministic approximations via ODEs are not accurate, (ii) the systems include species with low, medium, and high populations and thus the resulting state space of the stochastic process is prohibitively large to perform precise numerical analysis and existing reduction/approximation techniques are not sufficient (they are either too imprecise or do not provide sufficient reduction factors), and (iii) the system dynamics leads to bi-modal distributions and/or is affected by stiff reactions. 

These models thus represent perfect candidates for evaluating advanced approximation 
methods including various hybrid approaches~\cite{henzinger2010hybrid,Verena2013,HybridLNA2016}. 
Although these approaches can handle the models, they typically require tens of minutes or hours of computation time. Similarly 
simulation-based methods are very time consuming especially in case of very stiff CRN, represented by the viral infection model. We demonstrate that our approach provides accurate predications of the  system behaviour and is feasible even when performed manually by a human. 

Recall that the algorithm that builds the abstract model of the given CRN takes as input  two vectors representing the population discretisation and population bounds. We generally assume that these inputs are provided by users who have a priori knowledge about the system (e.g. in which orders the species population occurs) and 
that the inputs also reflect the level of details the users are interested in. In the following case studies, we, however, set the inputs only based on the rate orders of the reactions affecting the particular species (unless mentioned otherwise).

\subsection{Gene Expression Model}


The CRN underlying the gene expression model is described in Table~\ref{fig:gene}. As discussed in~\cite{Verena2013} and experimentally observed in~\cite{gandhi2011transcription}, the system oscillates  between two phases characterised by the  D\textsubscript{on}  state and the  D\textsubscript{off}  state, respectively. Biologists are interested in how the distribution of the  D\textsubscript{on}  and  D\textsubscript{off}  states is aligned with the distribution of RNA and proteins P, and how the correlation among the distributions  depends on the DNA  switching rates.

The state vector of the underlying CTMC is given as [P, RNA, D\textsubscript{off}, D\textsubscript{on}].  
We  use very relaxed bounds on the maximal populations, namely the bound 1000 for P and 100 for RNA. Note the DNA invariant   D\textsubscript{on}  +  D\textsubscript{off}  = 1. As in~\cite{Verena2013}, the initial state is given as [10,4,1,0].

We first consider the slow switching rates that lead to a more complicated dynamics including bimodal distributions. In order to demonstrate the refinement step and its effect on the accuracy of the model, we start with a very coarse abstraction. It distinguishes only the zero population and the non-zero populations
and thus it is not able to adequately capture the relationship between the DNA state and RNA/P population.
The pruned abstract model obtained using Algorithm 1 and 2 is depicted in Fig.~\ref{fig:gene1} (left).  The full one before pruning is shown in
\ifdefined\app Fig.~\ref{fig:gene1-full} in Appendix.
\else Fig.~6 \cite[Appendix]{techreport}.  \fi

The proposed analysis of the model identifies  the key trends in the system dynamic.
The red  transitions, representing iterations 1-3, capture the most probable paths in the system. The green component includes states with  DNA on (i.e.  D\textsubscript{on} = 1) where the system oscillates. The component is reached via the blue state with  D\textsubscript{off}  and no \mbox{RNAs}/P. The blue state is promptly reached from the initial state and then the system waits (roughly 100h according our rate abstraction)  for the next DNA activation. The oscillation is left via a deactivation in the iteration 4 (the blue dotted transition)\footnote{In Fig~\ref{fig:gene1}, the dotted transitions denote exit transitions representing the deactivations}. The estimation of the exit time computed using Algorithm 2 is also 100h. The deactivation  is then followed by fast red transitions leading to the blue state, where the system waits for the next activation. 
Therefore, we obtain an oscillation between the blue state and the green component, representing the expected oscillation between the D\textsubscript{on} and D\textsubscript{off} states. 

\begin{table}[t]
\setlength{\tabcolsep}{10pt}
\caption{Gene expression. For slow DNA switching, $r_1\!=\!r_2\!=\!0.05$. For fast DNA switching, $r_1\!=\!r_2\!=\!1$. The rates are in h$^{-1}$.}
\label{fig:gene}
\begin{tabular}{llll}
 D\textsubscript{off}  $\goes{r_1}$  D\textsubscript{on}  & D\textsubscript{on}  $\goes{r_2}$  D\textsubscript{off}  & D\textsubscript{on}  $\goes{10}$  D\textsubscript{on} + RNA  &  RNA  $\goes{1}$ $\emptyset$   \\
 RNA  $\goes{4}$ RNA + P  &  P $\goes{1}$ $\emptyset$ & P + D\textsubscript{off}  $\goes{0.0015}$  P + D\textsubscript{on} &    \\
\end{tabular}
\end{table}

\begin{figure}[t]
\includegraphics[width=\textwidth]{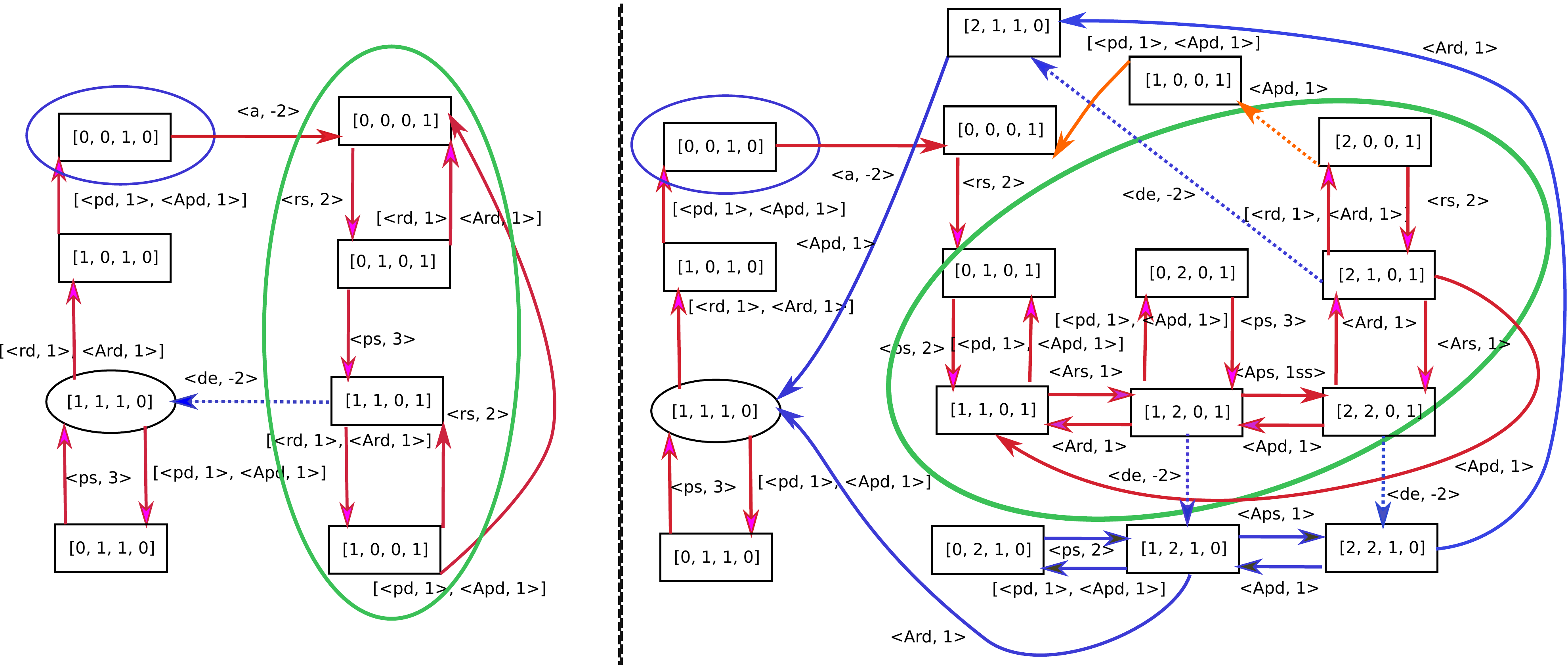}
\caption{Pruned abstraction for the gene expression model using the coarse population discretisation (left) and after the refinement (right). The state vector is [P, RNA, D\textsubscript{off}, D\textsubscript{on}].}
\label{fig:gene1}
\vspace{0.5em}
\end{figure}

As expected, this abstraction does not clearly predict the bimodal distribution on the RNA/P populations 
as the trivial population levels do not bear any information beside reaction enabledness.
In order to obtain a more accurate analysis of the system, we refine the population discretisation using a single level threshold for P and DNA, that is equal to 100 and 10, respectively (the rates in the CRN indicate that the population of P reaches higher values).

\new{Fig~\ref{fig:gene1} (right) depicts the pruned abstract model with the new discretisation (the full model is depicted in \ifdefined\app Fig.~\ref{fig:gene2} in Appendix.
\else Fig.~7 \cite[Appendix]{techreport}. \fi We again obtain the oscillation between the green component representing  DNA\textsubscript{on} states and the blue DNA\textsubscript{off} state. The states in the green component more accurately predicts that in the DNA\textsubscript{on} states
the populations of RNA and P are high and drop to zero only for short time periods.  The figure also shows orange transitions within the iteration 2 that extend the green component by two states. Note that the system promptly returns from these states back to the green component. After the deactivation in the iteration 4, the system takes (within the same iteration) the fast transitions (solid blue) leading to the blue component where system waits for another activation and where the mRNA/protein populations decrease.  The expected time spent in states on blue solid transitions is small and thus  we can reliably  predict the bimodal distribution of the mRNA/P populations and its correlation with the  DNA state. 
 The refined abstraction also reveals that the switching time from the DNA\textsubscript{on} mode to the DNA\textsubscript{off} mode is lower.  
 These predications are in accordance with the results obtained in~\cite{Verena2013}. See \ifdefined\app Fig.~\ref{fig:verena} in Appendix \else Fig.~8 \cite[Appendix]{techreport} \fi that is adopted from~\cite{Verena2013} and illustrates these results.}

To further test the accuracy of our approach, we consider the fast switching between the DNA states. We follow the study in~\cite{Verena2013} and increase the rates by two orders of magnitude. 
We use the refined population discretisation and  obtain a very similar abstraction as in Fig.~\ref{fig:gene1} (right). We again obtain the oscillation between the green component (DNA\textsubscript{on} states and nonzero RNA/protein populations) and the blue state (DNA\textsubscript{off} and zero RNA/protein populations). The only difference is in fact the transition rates corresponding to the activation and deactivation causing that the  switching rate between  the components is much faster. As a consequence, the system spends a longer period in the blue  transient states with D\textsubscript{off} and nonzero RNA/protein populations. The time spent in these states decreases the correlation between the DNA state and the RNA/protein populations as well as the bimodality in the population distribution. This is again in the accordance with~\cite{Verena2013}. 

To conclude this case study, we observe a very aligned agreement between the results obtained using our approach and results in~\cite{Verena2013} obtained via  advanced and time consuming numerical methods. We would like to emphasise that our abstraction and its solution is obtained within a fraction of a second while the numerical methods have to approximate solutions of equations describing  high-order conditional moments of the population distributions. As~\cite{Verena2013} does not report the runtime of the analysis and the implementation of their methods is not publicly available, we cannot  directly compare the time complexity.

\subsection{Goutsias's Model}
Goutsias's model illustrated in Table~\ref{fig:lambda}  is widely used for evaluation of various numerical and simulation based techniques. As showed e.g. in~\cite{goutsias2005quasiequilibrium}, the system has with a high probability the following transient behaviour.  In the first phase, the system switches with a high rate between the non-active DNA (denoted DNA) and the active DNA (DNA.D). During this phase the population of RNA, monomers (M) and dimers (D) gradually increase (with only negligible oscillations). After  around 15 minutes, the DNA is blocked (DNA.2D) and the population of RNA decreases while the population of M and D is relatively stable. After all RNA degrades (around another 15 minutes) the system 
switches to the third phase where the population of M and D slowly decreases. Further, there is a non-negligible  probability that the DNA is blocked at the beginning while the population of RNA is still small and the system promptly dies out.

\begin{table}[t]
\setlength{\tabcolsep}{5pt}
\caption{Goutsias' Model. The rates are in s$^{-1}$}
\label{fig:lambda}
\begin{tabular}{lll}
RNA  $\goes{0.043}$ RNA + M & M $\goes{7\times10^{-4}}$ $\emptyset$ &  RNA  $\goes{4\times10^{-3}}$ $\emptyset$  \\
DNA + D $\goes{0.002}$  DNA.D & DNA.D $\goes{0.48}$  DNA + D \\
DNA.D + D $\goes{2\times10^{-4}}$ DNA.2D & M+M  $\goes{0.083}$ D &  D  $\goes{0.5}$ M + M\\
 DNA.2D  $\goes{9\times10^{-12}}$ DNA.D + D & DNA.D $\goes{0.072}$  RNA + DNA.D 
\end{tabular}
\vspace{1em}
\end{table}

Although the system is quite suitable for the hybrid approaches (there is no strong bimodality and only a limited stiffness), the analysis still takes 10 to 50 minutes depending on the required precision~\cite{henzinger2010hybrid}. We demonstrate that our approach is able to accurately predict the main transient behaviour as well as the non-negligible probability that the system promptly dies out.

The state vector is given as [M, D, RNA, DNA, DNA.D, DNA.2D]  and the initial state is set to [2, 6, 0, 1, 0, 0] as in~\cite{henzinger2010hybrid}.  We start our analysis with a coarse population discretisation with a single threshold 100 for M and D and a single threshold 10 for RNA. We relax the bounds, in particular,  1000 for M and D, and 100 for RNA. Note that these numbers were selected solely based on the rate orders of the relevant reactions. 
Note the DNA invariant  DNA + DNA.D + DNA.2D = 1. 

\begin{figure}[t]
\centering
\includegraphics[width=1\textwidth]{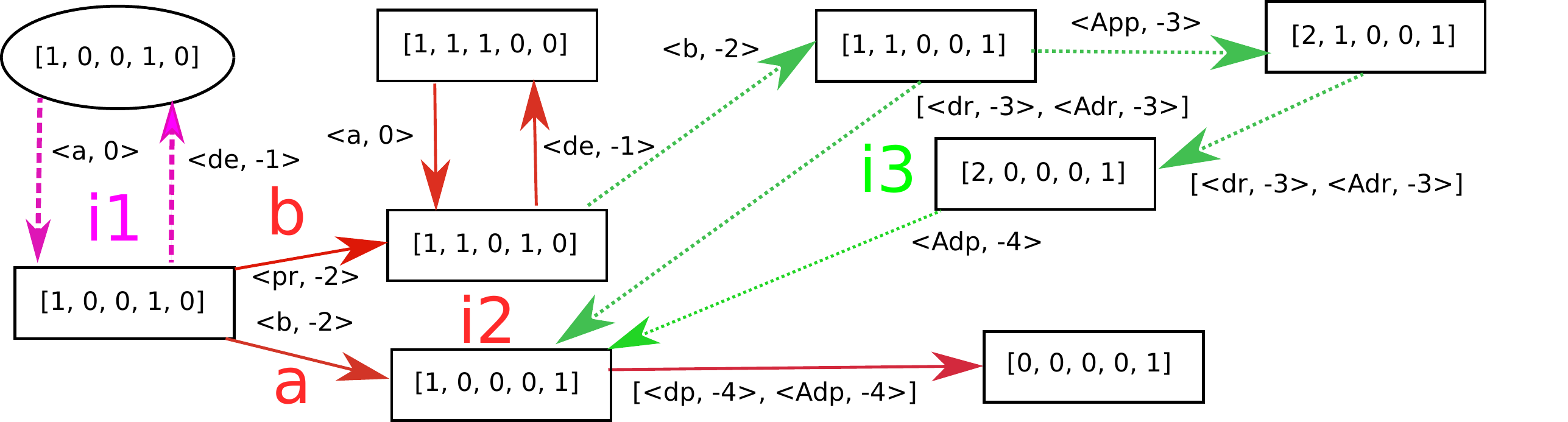}
\caption{Pruned abstraction for the Goutsias' model. The state vector is  [M+D, RNA, DNA, DNA.D, DNA.2D] }
\label{fig:lambda1}
\end{figure}

Fig.~\ref{fig:lambda1} illustrates the pruned abstract model we obtained (the full model is depicted in 
\ifdefined\app Fig.~\ref{fig:lambdafull} in Appendix.
\else Fig.~9 \cite[Appendix]{techreport}.  \fi
For a better visualisation, we merged the state components corresponding to M and D into one component with M+D. As there is the fast reversible dimerisation, the actual distributions between the population of M and D does not affect the transient behaviour we are interested~in. 

The analysis of the model shows the following transient behaviour. The purple dotted loop in the iteration \texttt{i1} represents (de-)activation of the DNA. The expected exit time of this loop is 100s. According to our abstraction, there are two options (with the same probability) to exit the loop: (1) the path \texttt{a} represents the DNA blocking followed  by the quick extinction and (2) the path \texttt{b}  corresponds to the production of $RNA$ and its followed by the red loop in the \text{i2} that again represents (de-)activation of the DNA. Note that according our abstraction, this loop contains states with the populations of  M/D as well as RNA up to 100 and 10, respectively.

The expected exit time of this loop is again 100s and there are two options how to leave the loop: 1) the path within the iteration $i3$ (taken with roughly 90\%) represents again the DNA blocking and it is followed by the extension of RNA and consequently  by the extension of M/D in about 1000s and 2) the path within the iteration 5 (shown in the full graph in \ifdefined\app Fig.~\ref{fig:lambdafull} in Appendix\else Fig.~9 \cite[Appendix]{techreport}\fi) taken with roughly 10\% represents the series of protein productions and leads to the states with a high number of proteins (above 100 in our population discretisation). Afterwards, there is again a series of DNA (de-)activations followed by the DNA blocking and the extinction of RNA. As before, this leads to the extinction of M/D in about 1000s.

Although this abstraction already shows the transient behaviour leading to the extinction in about 30 minutes, it introduces the following inaccuracy with respect to the  known behaviour: (1) the probability of the fast extinction is higher and (2) we do not observe the clear bell-shape pattern on the RNA (i.e. the level 2 for the RNA is not reached in the abstraction). 
As in the previous case study, the problem is that the population discretisation
is too coarse. It causes that the total rate of the DNA blocking (affected by the M/D population via the mass action kinetics) is too high in the states with the M/D population level 1. 
This can be directly seen in the interval CTMC representation where the rate spans many orders of magnitude, incurring too much imprecision.
The refinement of the M/D population discretisation eliminates the first inaccuracy. To obtain the clear bell-shape patter on RNA, one has to refine also the RNA  population discretisation. 

\subsection{Viral Infection}

The viral infection model described in Table~\ref{fig:viral} represents the most challenging system we consider.  It is highly stochastic, extremely stiff, with all species presenting high variance and some also very high molecular populations. Moreover, there is a bimodal distribution on the RNA population. As a consequence, the solution of the full CME, even using advanced reduction and aggregation techniques, is prohibitive due to state-space explosion and stochastic simulation are very time consuming. State-of-the-art hybrid approaches integrating the LNA and an adaptive population partitioning~\cite{HybridLNA2016} can handle this system but also need a very long execution time. For example, a transient analysis up to time $t = 50$ requires around 20 minutes and up to $t = 200$ more than an hour.

To evaluate the accuracy of our approach on this challenging  model, we also focus on the same transient analysis, namely, we are interested in the distribution of RNA at time $t=200$. The analysis in~\cite{HybridLNA2016} predicts a bimodal distribution where, the probability that RNA is zero in around 20\% and the remaining probability has Gaussian distribution with mean around 17 and the probability that there is more than 30 RNAs is close to zero. This is confirmed by simulation-based analysis in~\cite{goutsias2005quasiequilibrium} showing also the gradual growth of the RNA population. The simulation-based analysis in~\cite{srivastava2002stochastic}, however, estimates a lower probability (around 3\%) that RNA is 0 and higher mean of the remaining  Gaussian distribution (around 23).  Recall that obtaining accurate results using simulations is extremely time consuming due to very stiff reactions (a single simulation for $t=200$ takes around 20 seconds).  

In the final experiments, we analyse the distribution of RNA at time $t=200$ using our approach. The state vector is given as [P, RNA, DNA]  and we start with the concrete state [0, 1, 0]. To sufficiently reason about the RNA population and to handle the very high population of the proteins, we use the following population discretisation:  thresholds \{10,1000\} for P,  \{10,30\} for RNA, and \{10,100\} for DNA. As before, we use very relaxed bounds 10000, 100, and 1000 for P, RNA, and D, respectively. Note that we ignore the population of the virus V as it does not affect the dynamics of the other species. This simplification makes the visualisation of our approach more readable and has no effect on the complexity of the analysis.

\begin{table}[t]
\setlength{\tabcolsep}{10pt}
\caption{Viral Infection. The rates are day$^{-1}$}
\label{fig:viral}
\begin{tabular}{lll}
DNA + P  $\goes{7\times10^{-6}}$ V & DNA $\goes{0.025}$ DNA + R &  RNA  $\goes{0.25}$ $\emptyset$  \\
RNA $\goes{1}$ RNA + DNA & RNA $\goes{1000}$ RNA. + P & P $\goes{1.99} \emptyset$  
\end{tabular}
\vspace{0.5em}
\end{table}

\begin{figure}[t]
	\centering
	\includegraphics[width=1\textwidth]{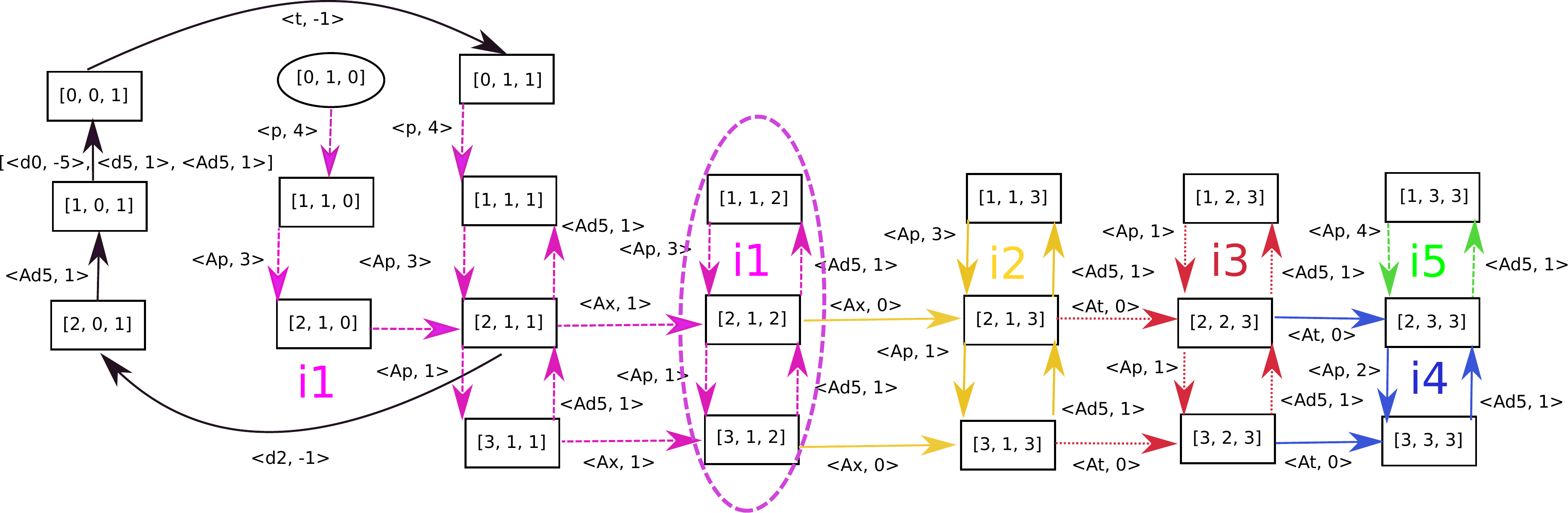}
	\caption{Pruned abstraction for the viral infection model. The state vector is [P, RNA, DNA].}
	\label{fig:viral-pruned}
	\vspace{1em}
\end{figure}

Fig.~\ref{fig:viral-pruned} illustrates the obtained abstract model enabling the following transient analysis (the full model is depicted in 
\ifdefined\app Fig.~\ref{fig:viral1} in Appendix.
\else Fig.~10 \cite[Appendix]{techreport}.  \fi
In a few days the system reaches from the initial state the  loop (depicted by the purple dashed ellipse) within the iteration~\textit{i1}. The loop includes states where RNA has level~1, DNA has level~2 and P oscillates between the levels 2 and 3. Before entering the loop, there is a non-negligible probability (orders of percent) that the RNA drops to 0 via the full black branch that returns to  transient part of the loop in~\textit{i1}. In this branch the system can also die out (not shown in this figure, see the full model) with probability in the order of tenths of percent.

The average exit time of the loop in \textit{i1} is in the order of 10 days and the system goes to the yellow loop within the iteration  \textit{i2}, where the DNA level  is increased to 3 (RNA level is unchanged and  and P again oscillates between the levels 2 and~3). The average exit time of the loop in \textit{i2} is again in the order of 10 days and systems goes to the dotted red loop within iteration \textit{i3}. The transition represents the sequence of RNA synthesis that leads to RNA level 2. P oscillates as before. Finally,  the system leaves the loop in \textit{i3} (this takes another dozen days) and reaches RNA level 3 in iterations \textit{i4} and \textit{i5} where the DNA level remains at the level 3 and P oscillates. The iteration \textit{i4} and \textit{i5} thus roughly correspond to the examined transient time $t=200$.

The analysis clearly demonstrates that our approach leads to the behaviour that is well aligned with the previous experiments. We observed growth of the RNA population with a non-negligible probability of its extinction. The concrete quantities (i.e. the probability of the extinction and the mean RNA population) are closer to the analysis in~\cite{srivastava2002stochastic}. The quantities are indeed affected by the population discretisation and can be further refined. We would like to emphasise that in contrast to the methods presented in~\cite{srivastava2002stochastic,goutsias2005quasiequilibrium,HybridLNA2016} requiring  hours  of intensive numerical computation, our approach can be done even manually on the paper.

%

\bibliographystyle{splncs04}
\bibliography{bib,bib2,bib3,new_bib}

\ifdefined\app

 \newpage
 \appendix

\section{Additional Figures}

\begin{figure}[h!]
\centering
\includegraphics[width=0.7\textwidth]{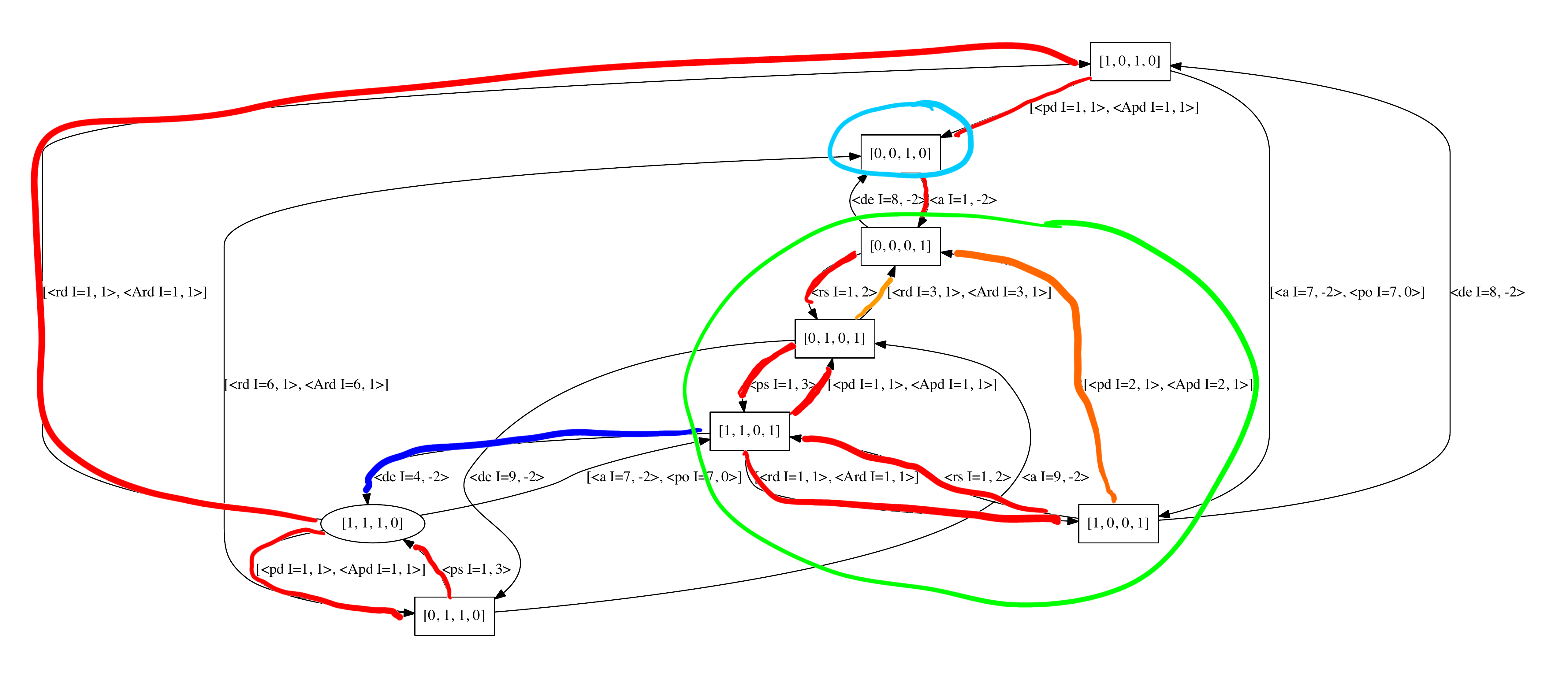}
\caption{The course abstraction for the gene expression model.}
\label{fig:gene1-full}
\end{figure}

\begin{figure}[h!]
\includegraphics[width=\textwidth]{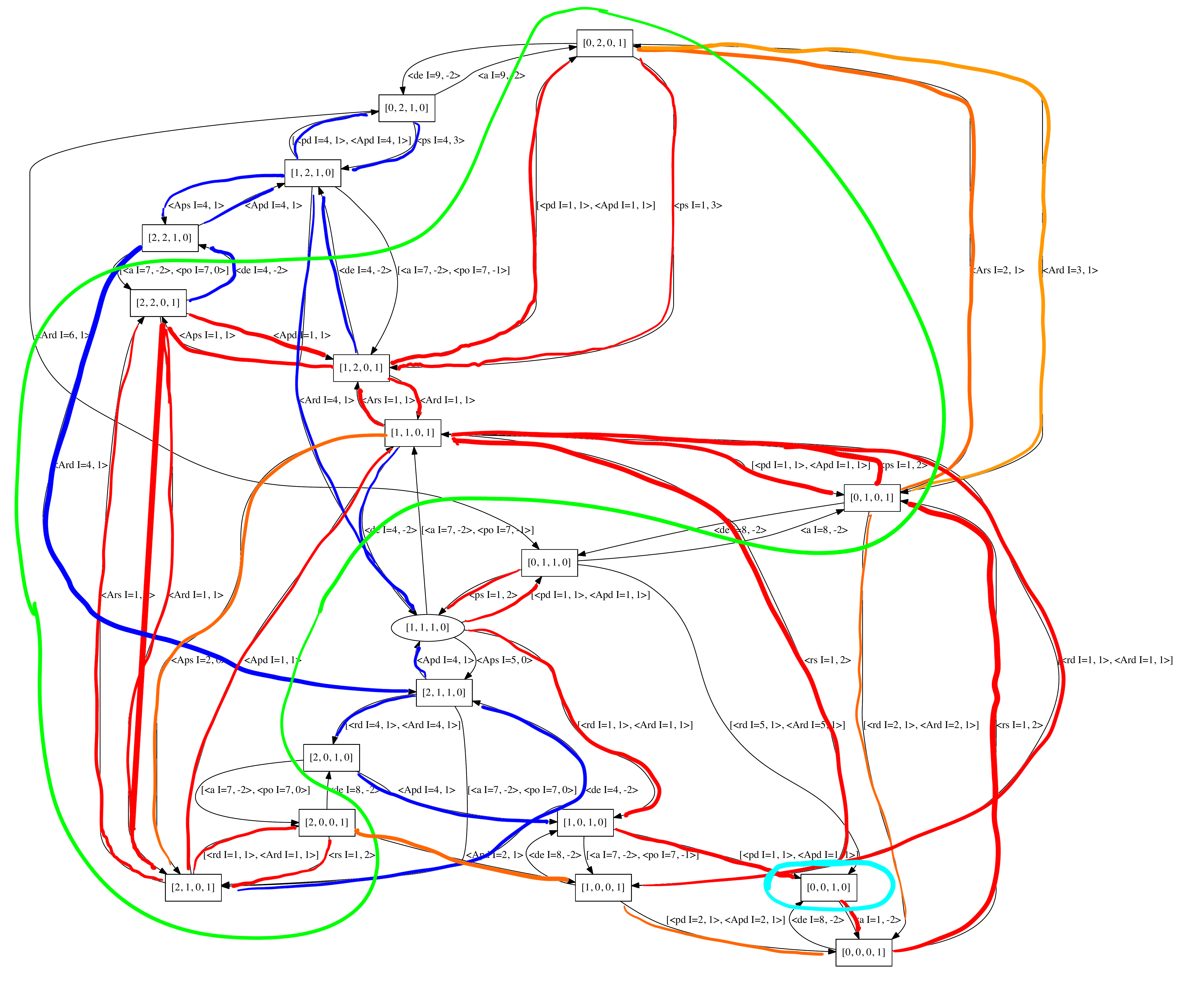}
\caption{The refined abstraction for the gene expression model.}
\label{fig:gene2}
\end{figure}

\begin{figure}[h!]
\centering
\includegraphics[width=0.6\textwidth,angle=270]{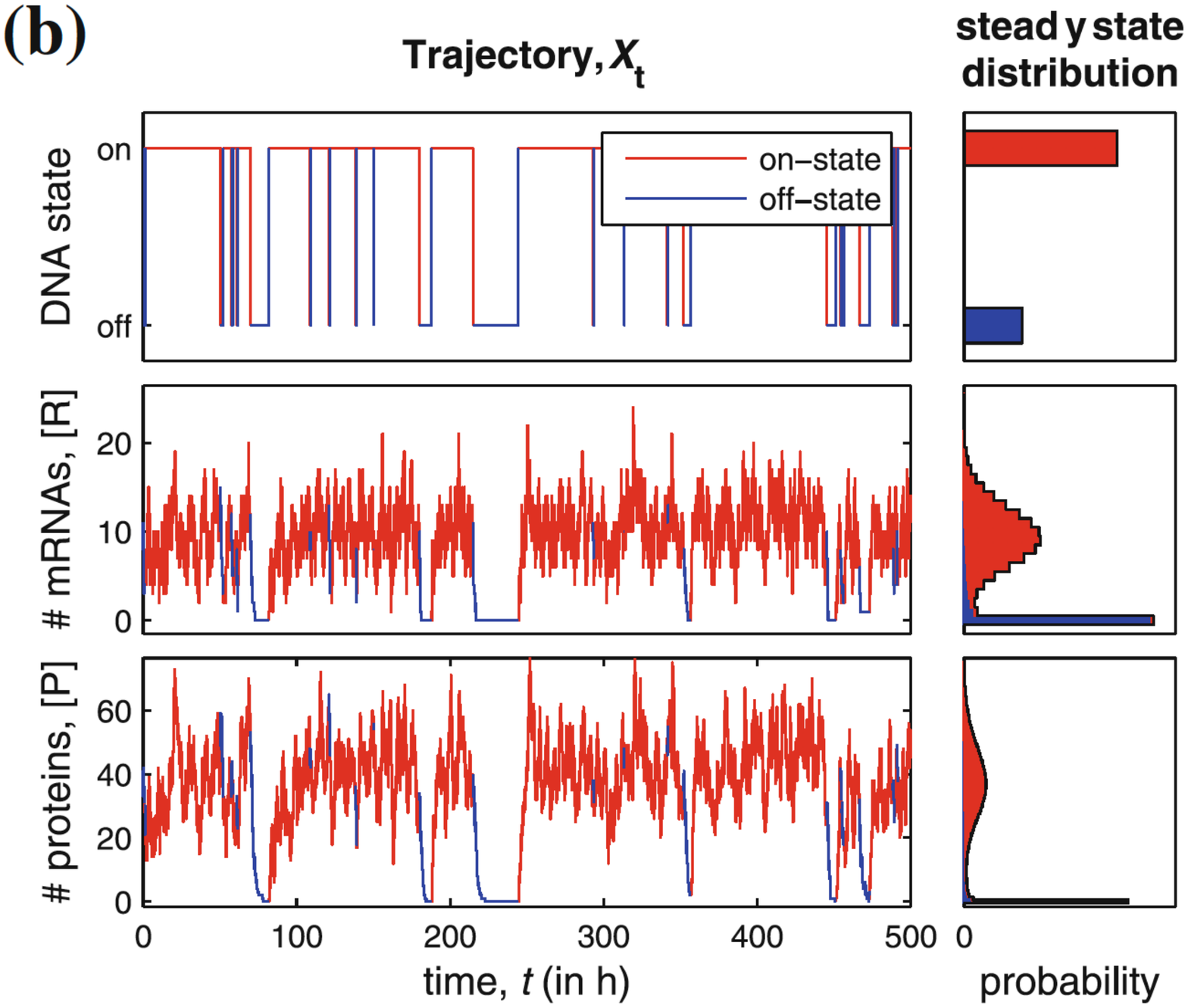}
\caption{The gene expression model: illustration of the bimodal distribution of the mRNA/P populations and its correlation with the  DNA state. Adopted from~\cite{Verena2013}. }
\label{fig:verena}
\end{figure}

\begin{figure}[h!]
\includegraphics[width=\textwidth]{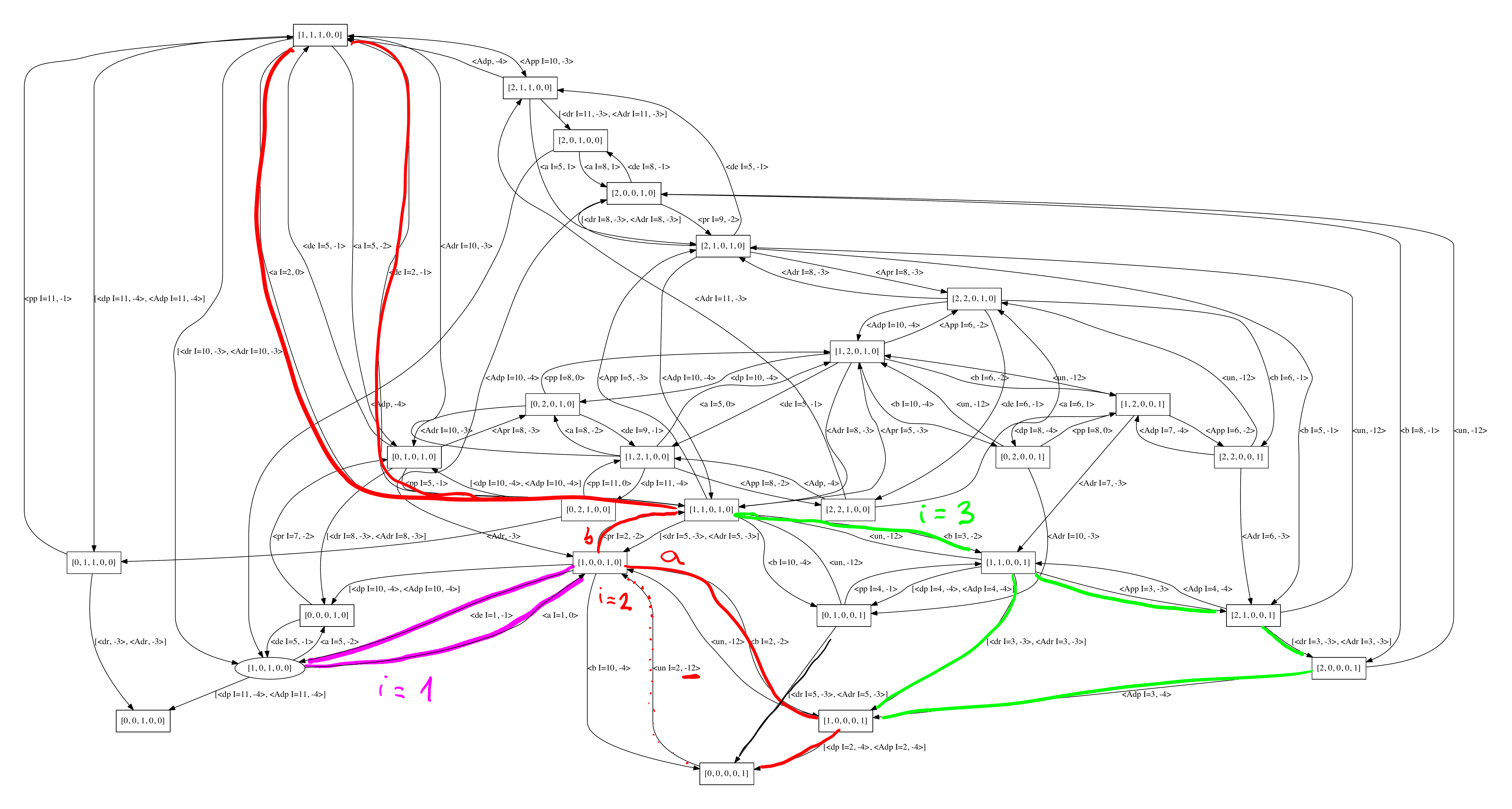}

\caption{The abstraction for the Goutsias' model.}
\label{fig:lambdafull}
\end{figure}

\begin{figure}[h!]
\includegraphics[width=\textwidth]{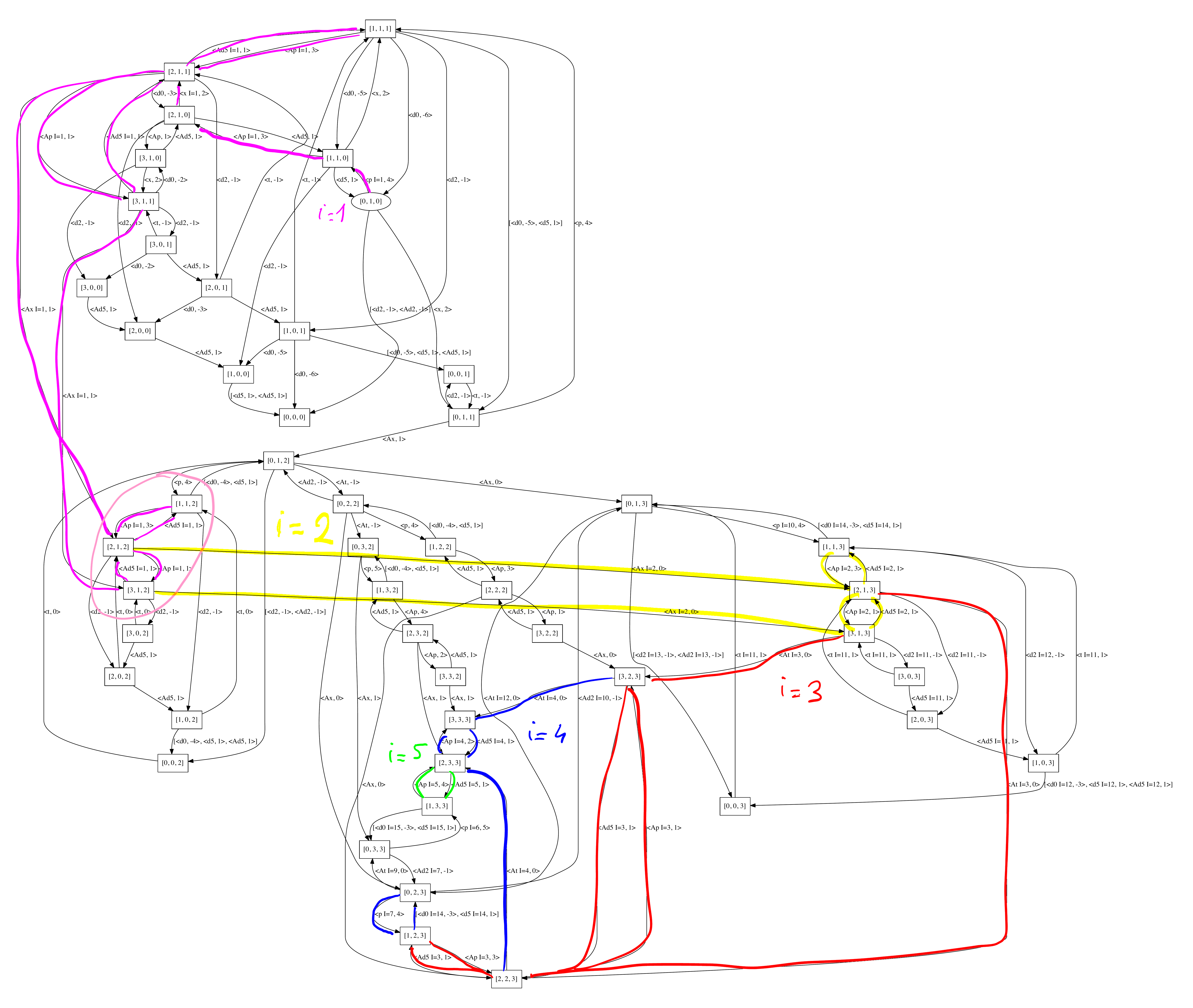}
\caption{The abstraction for the viral infection model.}
\label{fig:viral1}
\end{figure}

\clearpage

\section{Continuous-time Markov chains}

\newcommand{\model}{\ensuremath{\mathcal{C}}}
\newcommand{\distr}{\pi}
\newcommand{\tmat}{\mathbf{R}}
\newcommand{\states}{S}
\newcommand{\state}{s}
\newcommand{\pmat}{\mathbf{P}}
\newcommand{\CTMC}{\ensuremath{\mathcal{C}=(S,\pi_0, \mathbf{R})}}
\newcommand{\pp}{\ensuremath{\gamma}}
\newcommand{\ddistr}{\tau}
\newcommand{\Path}{\mbox{Path}}
\newcommand{\setofruns}{\mathit{Runs}}
\newcommand{\sigmafield}{\mathcal F}
\newcommand{\probm}{\mathbb P}

Here we recall the standard definition of continuous-time Markov chains (CTMC).

We denote the set of non-negative real numbers by$ \mathbb{R}_{\geq 0}$.
A \emph{probability distribution} on a finite or countably infinite set $X$ is a mapping $\delta: X \to [0,1]$, such that $\sum_{x\in X} \delta(x) = 1$.
The set of all probability distributions on $X$ is denoted by $\mathcal D(X)$.
A probability distribution $\delta$ is Dirac if there exists $x\in X$ with $\delta(x)=1$.

A~CTMC is a tuple $\CTMC$~where:
	\begin{itemize}
		\item $\states$ is a finite or countably infinite set of {\em states};
		\item $\distr_{0}\in\mathcal D(\states)$ is the {\em initial distribution} (possibly Dirac);
		\item $\mathbf{R}: S \times S \rightarrow \mathbb{R}_{\geq 0}$ is the {\em rate matrix}.
	\end{itemize}

A run $\omega$ of $\mathcal{C}$ is an infinite sequence
$\omega = s_0t_0s_1t_1\ldots$, where for all $i$, $s_i \in S$ and $t_{i} \in \mathbb{R}_{\geq 0}$ is the time spent in state $s_i$.
A run of a CTMC $\model$ is initiated in some state $s_0$, 
which is chosen randomly according to $\distr_0$. 
In the 
current state $s_i$, the next state $s_{i+1}$ is selected randomly as follows.
A transition between states $s,s'\in S$ can occur only if $\mathbf{R}(s,s')>0$ and, in that case, the probability of triggering the transition within
time $t$ is $1-e^{-t\cdot\mathbf{R}(s,s')}$.
Consequently, the time spent in state $s$, before a transition is triggered, is exponentially distributed with {\em exit rate} $E(s)=\sum_{s'\in S}\mathbf{R}(s,s')$,
and when the transition occurs the probability of moving to state $s'$ is given by $P(s,s')=\frac{\mathbf{R}(s,s')}{E(s)}$.

We use $\setofruns_{\model}$
to denote the set of all runs of $\model$.
Now we define a probability space 
$(\setofruns_{\model},\sigmafield_{\model},\probm_{\model})$
over the runs of~$\model$. A \emph{template} is a finite sequence of the form 
$B = s_0\,I_0\,s_1\,I_1\cdots s_{n+1}$ such that
$n \geq 0$ and $I_i$ is an interval in~$\mathcal R_{\geq 0}$
for every $0 \leq i \leq n$. Each such~$B$ determines the
corresponding \emph{cylinder} $\setofruns(B) \subseteq \setofruns_{\model}$
consisting of all runs of the form $\hat{s}_0\,t_0\,\hat{s}_1\,t_1\cdots$, where
$\hat{s}_i = s_i$ for all \mbox{$0 \leq i \leq n{+}1$}, and $t_i \in I_i$ for
all $0 \leq i \leq n$. The
$\sigma$-field $\sigmafield_{\model}$ is the Borel $\sigma$-field generated by all
cylinders. For each template $B = s_0\,I_0\,s_1\,I_1\cdots s_{n+1}$, the probability $\probm_{\model}(\setofruns(B))$ is 
defined as follows:
\[
\distr_0(s_0) \cdot 
\prod_{i=0}^{n} P(s_i,s_{i+1}) \cdot (e^{-E(s_i)\cdot \inf I_i}-e^{-E(s_i)\cdot \sup I_i})
\]
Then, $\probm_{\model}$ is extended to $\sigmafield_{\model}$ in the unique way by applying 
Carath\'eodory extension theorem (see, e.g., \cite{baier2008principles}).

\fi
 
\end{document}